\documentclass[aps,prb,twocolumn,superscriptaddress,longbibliography,nofootinbib,floatfix]{revtex4-2}
\usepackage{amsmath,amssymb,bm}
\usepackage{graphicx}
\usepackage{xcolor}
\usepackage{hyperref}
\hypersetup{colorlinks=true,linkcolor=blue!60!black,citecolor=blue!60!black,urlcolor=blue!60!black}

\newcommand{\zOne}{0.511(2)}
\newcommand{\GcOne}{2.5225(14)}
\newcommand{\zTwo}{0.661(2)}
\newcommand{\GcTwo}{2.0617(6)}
\newcommand{\zThree}{0.831(15)}
\newcommand{\GcThree}{1.7629(9)}
\newcommand{\Lmax}{128}

\begin{document}

\title{Crossover duality for the quantum long-range Ising chain:\\
perturbation theory around the short-range endpoint and a DMRG test}

\author{Junchen Rong}
\affiliation{CPHT, CNRS, \'Ecole Polytechnique, Institut Polytechnique de Paris, Palaiseau, France}

\date{\today}

\begin{abstract}
We propose a weakly coupled description of the long-range transverse-field Ising chain near its short-range crossover at $s_*=7/4$, where the couplings decay as $r^{-1-s}$. The Ising conformal field theory is coupled to an instantaneous Gaussian field through $\sigma\chi$, extending the classical long-range Ising duality to quantum chains. The equal-time interaction generates a stress-tensor deformation that cause the velocity to run and renormalizes the dynamical critical exponent $z$. Conformal perturbation theory gives $z=1-0.720\,\delta+O(\delta^2)$, $1/\nu=(1+z)/2+O(\delta^2)$, and $\omega=\delta+O(\delta^2)$, with $\delta=s_*-s$. DMRG calculations on periodic chains of up to $\Lmax$ sites provide a consistency test using theoretical inputs for $1/\nu$ and $\omega$. After accounting for slow finite-size corrections, static and dynamical estimates of $z$ agree, supporting the crossover expansion at $s=1.5$ and a two-sided Pad\'e interpolation at $s=1$, $1.25$, and $1.5$.
\end{abstract}

\maketitle

\section{Introduction}
\label{sec:intro}

Quantum spin chains with algebraically decaying interactions are no longer a theoretical abstraction. Trapped-ion quantum simulators realize the transverse-field Ising model with spin--spin couplings $J_{ij}\propto|i-j|^{-\alpha}$ mediated by the collective motional modes, and the exponent $\alpha$ can in principle be tuned continuously over $0<\alpha<3$ by the detuning of the Raman beams from the phonon spectrum~\cite{PorrasCirac2004,Islam2011,Britton2012,MonroeRMP}. In our notation $\alpha=1+s$. The critical line of the chain has a mean-field part, $\alpha<5/3$, and a non-mean-field part, $5/3<\alpha<11/4$ ($2/3<s<7/4$), in which the critical exponents depend continuously on $s$ and are not known in closed form; beyond $\alpha=11/4$ the transition is in the short-range Ising class. Most experiments to date, including the measurements of critical behavior---the onset of the ferromagnetic transition~\cite{Islam2011,Islam2013}, Kibble--Zurek scaling across it in chains of up to $61$ ions~\cite{Li2023}, and nonequilibrium critical scaling after a quench~\cite{De2025}---operate at $\alpha\approx1$, on the mean-field side, because a larger $\alpha$ requires detuning farther from the motional modes at the cost of coupling strength and coherence; exponents up to $\alpha\approx1.75$ have been reached in short chains~\cite{Jurcevic2014}. The non-mean-field window is therefore at the edge of what present ion-trap experiments access, and it is not reached by the other long-range platforms, whose exponents are fixed by the interaction---van der Waals Rydberg arrays ($\alpha=6$) and dipolar systems ($\alpha=3$) both lie on the short-range side of the crossover for a chain (for reviews see Refs.~\cite{DefenuRMP,MonroeRMP}). Theoretical control of the window $2/3<s<7/4$ is thus the input that will be needed when experiments move there, and it is the subject of this paper.

The theoretical situation along this line is the following. Near $s=2/3$ the order parameter is almost Gaussian and a fractional $\phi^4$ expansion in $\epsilon=3s/2-1$ applies~\cite{Dutta2001,Maghrebi2016,Defenu2017,Benedetti2024}. Near $s_*=7/4$ the critical theory is almost the short-range (SR) Ising conformal field theory (CFT), but the fractional $\phi^4$ description has no small parameter there. For the \emph{classical} long-range Ising model the analogous problem was solved by Behan, Rastelli, Rychkov, and Zan~\cite{BRRZ1,BRRZ2}, who proposed an infrared duality: the long-range fixed point can be reached either from a nonlocal Gaussian field perturbed by $\phi^4$, or from the SR Ising CFT tensored with a generalized free field and perturbed by the bilinear $\sigma\chi$. The second description is weakly coupled near the crossover and resolves otherwise puzzling features of it, such as the persistence of the shadow relation and the recombination of the stress-tensor multiplet.

The classical chain with exchange $J(r)\propto r^{-1-s}$ orders at finite temperature for $0<s<1$~\cite{Dyson1969} and at $s=1$~\cite{FrohlichSpencer1982}, where the spontaneous magnetization jumps at the transition~\cite{Thouless1969,Aizenman1988}. The conformal description of the classical long-range Ising model~\cite{Paulos2016} has been tested in one dimension by Monte Carlo simulations~\cite{AntunesMatos2026}, perturbative expansions~\cite{Benedetti1d2025,Benedetti1d2026,Zheng2026}, and bootstrap sum rules~\cite{GhoshPaulos2026}, and in two and three dimensions by conformal bootstrap calculations~\cite{Behan2019,BehanBootstrap2024}. The quantum chain presents the same opportunity and a new obstruction. Its long-range interaction is instantaneous in imaginary time, so the auxiliary field that localizes it must have the correlator $\langle\chi(\tau,x)\chi(0,0)\rangle\propto\delta(\tau)/|x|^{1+s}$. It is Gaussian, but it is not an isotropic generalized free field. The temporal delta function selects equal-time configurations of the Ising correlators, which changes the beta-function integral and, more importantly, generates a running space-time anisotropy. The latter is the origin of a dynamical exponent $z\neq1$, which has no analog in the classical problem. The first purpose of this paper is to carry out this anisotropic conformal perturbation theory to leading nontrivial order and to extract $z$, $\nu$, and the correction-to-scaling exponent $\omega$ as expansions in $\delta=s_*-s$ (Sec.~\ref{sec:theory}).

The second purpose is to test the result numerically in the middle of the critical line, where neither weak-coupling expansion is \emph{a priori} controlled. Quantum Monte Carlo (QMC) studies of the chain~\cite{Koziol2021,Shiratani2024} report a systematic difficulty near $s_*$: finite-size estimates of $z$ depend strongly on whether corrections to scaling are modeled as a power law or as a logarithm, with the two extrapolations differing by up to $0.07$ as $s\to s_*$. The crossover expansion explains this: the leading irrelevant eigenvalue vanishes linearly at the crossover, $\omega=\delta+O(\delta^2)$, so finite-size corrections decay as $L^{-\delta}$ and become arbitrarily slow as $s\to s_*$. We therefore perform density-matrix renormalization group (DMRG)~\cite{White1992,TeNPy} calculations on periodic chains with the periodic images of the interaction summed exactly, and we extract $z$ in two independent ways: from the finite-size gap and, without any reference to dynamics, from the static spin correlator through the exact relation $2\Delta_\sigma=1+z-s$ that follows from the nonrenormalization of the long-range kernel. The two determinations disagree at accessible sizes near the crossover, as they should, and converge to a common value once the $L^{-\omega}$ corrections predicted by the expansion are taken into account (Sec.~\ref{sec:dmrg}). We emphasize the logic of this comparison: the exponents $\omega$ and $1/\nu$ that control the finite-size analysis are \emph{inputs} from the theory, not fitted quantities, so the numerics do not provide an independent measurement of $z$; they test whether the lattice data are consistent with the crossover expansion when its own corrections to scaling are used. With this input, the value of $z$ at $s=1.5$ agrees with the crossover expansion, and the values at all three $s$ agree with a two-sided Pad\'e interpolation between the two weakly coupled ends.

Throughout we distinguish three statements whose logical status is different. The Sak crossover criterion and the localization of the bilocal interaction are from scaling analysis. The existence of the weakly coupled fixed point follows perturbatively once the proposed $\text{Ising}_2\otimes\chi$ theory is assumed renormalizable with the counterterms identified below. The identification of this fixed point with the quantum long-range Ising fixed point is an infrared-duality proposal, which is what the numerics of Sec.~\ref{sec:dmrg} test.

\section{Duality and perturbation theory around the crossover}
\label{sec:theory}

\subsection{The quantum long-range Ising chain and its continuum description}
\label{sec:ends}

We study the ferromagnetic transverse-field chain
\begin{equation}
H=-J\sum_{i<j}\frac{Z_iZ_j}{|i-j|^{1+s}}-\Gamma\sum_i X_i,\qquad s>0,
\label{eq:H}
\end{equation}
with Pauli matrices $Z_i,X_i$. Its imaginary-time path integral is local in $\tau$, because both the interaction and the transverse field act within a time slice, and nonlocal in $x$. The Fourier transform of the coupling, $\hat J(k)=\sum_{r\neq0}|r|^{-1-s}e^{ikr}=2\,\mathrm{Re}\,\mathrm{Li}_{1+s}(e^{ik})$, is nonanalytic at $k=0$,
\begin{equation}
\hat J(k)=2\zeta(1+s)-c_s|k|^s-\zeta(s-1)k^2+O(k^4),
\label{eq:Jk}
\end{equation}
with $c_s=2|\Gamma(-s)\cos\tfrac{\pi s}{2}|>0$ for all $0<s<2$. The nonanalytic term is the leading $k$ dependence, and since the interaction is instantaneous $\hat J$ carries no frequency dependence; the $\omega^2$ term below comes from the transverse field alone. The Hubbard--Stratonovich order parameter $\phi$ is therefore governed, at the critical field, by
\begin{align}
S_\phi={}&\frac12\int_{\omega,k}\big(\omega^2+c_s|k|^s+c_2k^2+r\big)|\phi(\omega,k)|^2\nonumber\\
&+\frac{u}{4!}\int d\tau\,dx\,\phi^4 .
\label{eq:Sphi}
\end{align}
This is a fractional Lifshitz theory: the Gaussian part is invariant under $x\to bx$, $\tau\to b^{z_0}\tau$ with $z_0=s/2$, under which $\phi$ has dimension $\Delta^{(0)}_\phi=\tfrac12(1+z_0-s)=\tfrac12-\tfrac s4$ and the quartic coupling has eigenvalue $y_u=1+z_0-4\Delta^{(0)}_\phi=\tfrac{3s}2-1$. At the Gaussian fixed point the analytic $k^2$ term is irrelevant with correction exponent $2-s$. In the mean-field regime $s<2/3$, the quartic coupling has the smaller correction exponent $1-3s/2$. On the interacting side near $s=2/3$, the quartic correction exponent is instead $\omega=\epsilon+O(\epsilon^2)$; this endpoint expansion does not determine where other corrections become dominant farther along the critical line. For $s<2/3$ the fixed point is Gaussian, with $z=s/2$, $\nu=1/s$, and $\Delta_\phi=\tfrac12-\tfrac s4$. For $s=\tfrac23+O(\epsilon)$, $\epsilon=\tfrac{3s}{2}-1$, a Wilson--Fisher-like long-range fixed point (LRFP) exists at $u_*=O(\epsilon)$. Because the $\phi^4$ vertex is local in both $\tau$ and $x$, it renormalizes the analytic $\omega^2$ term but not the nonanalytic $|k|^s$ term; the resulting temporal wave-function renormalization is what moves $z$ away from $s/2$, first at two loops~\cite{Dutta2001,Maghrebi2016,Defenu2017,Benedetti2024}:
\begin{equation}
z=\frac s2+0.0311\,\epsilon^2+O(\epsilon^3).
\label{eq:zMF}
\end{equation}
The numerical coefficient quoted here inherits the approximately one-percent precision of the two-loop integral in Ref.~\cite{Benedetti2024}.
The LRFP is scale invariant with anisotropic scaling but has no reason to be conformal in any sense: in $1+1$ dimensions rotations are already broken by the Lifshitz dispersion, and the spatial nonlocality excludes a local stress tensor. Energy and momentum are, of course, conserved.

The description \eqref{eq:Sphi} is weakly coupled only near $s=2/3$. The rest of this section is concerned with the opposite end of the critical line.

\subsection{Sak's argument and the short-range crossover}
\label{sec:sak}

For large $s$ the chain is in the universality class of the nearest-neighbor model, whose critical point is the two-dimensional Ising CFT with $z=1$, $\Delta_\sigma=1/8$, i.e.\ $\eta_{\rm SR}=1/4$. Where does the long-range (LR) behavior give way to the short-range (SR) one? Fisher, Ma, and Nickel~\cite{FMN1972} compared the two quadratic terms in \eqref{eq:Sphi} and placed the crossover at $s=2$. Sak~\cite{Sak1973} pointed out that this ignores the renormalization of the analytic term at the SR fixed point, where $G^{-1}_{\rm SR}(0,k)\sim|k|^{2-\eta_{\rm SR}}$; the nonanalytic term $|k|^s$ competes with $|k|^{2-\eta_{\rm SR}}$, not with $k^2$, so the crossover is at
\begin{equation}
s_*=2-\eta_{\rm SR}=\frac74 .
\end{equation}
In CFT language the same statement reads as follows. At the SR quantum critical point the long-range tail of the interaction is the bilocal, equal-time perturbation
\begin{equation}
S_{\rm LR}=-\frac\lambda2\int d\tau\,dx\,dy\,\frac{\sigma(\tau,x)\sigma(\tau,y)}{|x-y|^{1+s}} ,
\label{eq:SLR}
\end{equation}
whose RG eigenvalue at $z=1$ is $y_\lambda=3-(1+s)-2\Delta_\sigma=2-s-\eta_{\rm SR}$: three powers of length from the measure, $-(1+s)$ from the kernel, $-2\Delta_\sigma$ from the two spins. It is relevant for $s<s_*$ and irrelevant for $s>s_*$. We write
\begin{equation}
\delta\equiv s_*-s,\qquad y_\lambda=\delta .
\end{equation}

Sak's argument has a second part, which is the one we will use most. Integrating out short-distance modes generates corrections to the quadratic action that are analytic in $k$; they can never produce or modify a $|k|^s$ term. The coefficient $c_s$ is therefore an RG invariant throughout the LR regime, and demanding that the term $\int d\tau\,dk\,|k|^s|\phi|^2$ have dimension zero under $x\to bx$, $\tau\to b^z\tau$, $\phi\to b^{-\Delta_\sigma}\phi$ gives
\begin{equation}
2\Delta_\sigma=1+z-s\qquad(2/3<s<s_*),
\label{eq:shadow}
\end{equation}
the quantum version of the classical statement $\eta=2-s$~\cite{FMN1972,Sak1973,Luijten2002,Angelini2014}. Two consequences are worth stating at once. First, the LR critical line has only two independent exponents, say $z$ and $\nu$, rather than three. Second, continuity of $\Delta_\sigma$ at the crossover requires $z\to1$ as $s\to s_*^-$; the way in which $z$ approaches $1$ is the first thing that the dual description computes. The exactness of \eqref{eq:shadow} has been debated in the classical setting~\cite{Honkonen1989,Luijten2002}; the dual description of the next subsection makes it manifest, as it does in the classical case~\cite{BRRZ1,BRRZ2}.

\subsection{Infrared duality: the classical case and its quantum analogue}
\label{sec:duality}

\paragraph*{The classical chain.}
For the classical chain, cluster simulations resolve mean-field finite-size scaling for $0<s<1/2$~\cite{LuijtenBlote1997} and give estimates of the varying exponents for $1/2<s<1$~\cite{Uzelac2001}. 
See~\cite{Pagni2026,AYH1970,Kosterlitz1976,Tower2026,LuijtenMessingfeld2001,Brezin2014} for papers that studied various aspects of this model.
\paragraph*{The classical duality.}
For the $d$-dimensional classical long-range Ising model, $J(r)\sim r^{-d-s}$, Behan, Rastelli, Rychkov, and Zan~\cite{BRRZ1,BRRZ2} proposed that near $s_*=2-\eta_{\rm SR}$ the LRFP is reached from a second ultraviolet starting point: the SR Ising CFT tensored with a generalized free field (GFF) $\chi$ of dimension $\Delta_\chi=(d+s)/2$, i.e.\ $\langle\chi(x)\chi(0)\rangle=|x|^{-d-s}$, perturbed by the bilinear
\begin{equation}
S=S_{\rm SRFP}+S_\chi-g\int d^dx\,\sigma\chi .
\label{eq:Sclass}
\end{equation}
Integrating out $\chi$ reproduces the bilocal interaction, so \eqref{eq:Sclass} is simply the Hubbard--Stratonovich form of the SR theory plus its long-range tail; the point is that $\chi$ carries the nonlocality and $g$ is a local, weakly relevant coupling, $\Delta_{\sigma\chi}=d-\delta/2$. Since $\chi\to-\chi$ maps $g\to-g$, the beta function is odd, $\beta_g=-\tfrac\delta2g+\beta_3g^3+\dots$, and $\beta_3>0$ produces an infrared fixed point at $g_*^2=\delta/2\beta_3$. The claim is that this fixed point is the LRFP of the nonlocal $\phi^4$ description, so that the two theories are ultraviolet-distinct descriptions of one infrared fixed point, each weakly coupled at one end of the critical line (``infrared duality''). The evidence is structural. The $\chi$ kinetic term is nonlocal and cannot be renormalized by local counterterms, so $\Delta_\chi=(d+s)/2$ exactly; the equation of motion $|k|^{-s}\chi\propto g\sigma$ then gives $\Delta_\chi=\Delta_\sigma+s$, hence $\Delta_\sigma=(d-s)/2$ and the shadow relation $\Delta_\sigma+\Delta_\chi=d$---Sak's nonrenormalization theorem, now manifest to all orders. On the $\phi^4$ side the equation of motion $|k|^s\phi\propto u\phi^3$ gives $\Delta_{\phi^3}=\Delta_\phi+s$, so the dictionary is $\phi\leftrightarrow\sigma$, $\phi^2\leftrightarrow\varepsilon$, $\phi^3\leftrightarrow\chi$. Finally, the LRFP has no local stress tensor; in the dual description this appears as multiplet recombination, $\partial^\mu T_{\mu\nu}=b(g)V_\nu$ with $V_\nu\propto\Delta_\sigma\,\sigma\partial_\nu\chi-\Delta_\chi\,\chi\partial_\nu\sigma$ a vector primary of the decoupled theory~\cite{BRRZ2,Paulos2016}. In $d=2$ the cubic coefficient is a regulated integral of the exact Ising four-spin function, $\beta_3=1.2684040(5)$~\cite{BRRZ1}, and the energy operator acquires no anomalous dimension at $O(g^2)$.

\begin{figure}[t]
\centering
\includegraphics[width=\columnwidth]{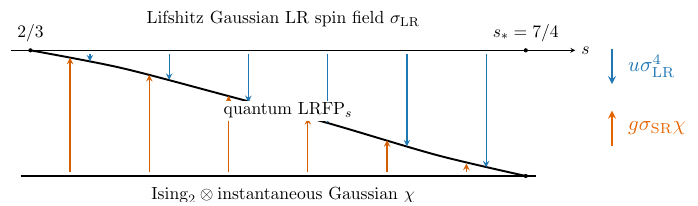}
\caption{Two complementary weak-coupling descriptions of the quantum long-range fixed-point line. The upper Lifshitz Gaussian theory flows under $u\sigma^4_{\rm LR}$ and controls the neighborhood of $s=2/3$. The lower product theory flows under $g\sigma_{\rm SR}\chi$ and controls the neighborhood of $s_*=7/4$. The vertical placement is schematic; the middle curve denotes the proposed common long-range fixed point.}
\label{fig:duality}
\end{figure}

\paragraph*{The quantum analogue.}
The lattice interaction in \eqref{eq:H} is instantaneous. We define the auxiliary Gaussian field by the covariance
\begin{equation}
\langle\chi(\tau,x)\chi(0,0)\rangle_0=N_\chi\,\frac{\delta(\tau)}{|x|^{1+s}},\qquad N_\chi>0,
\label{eq:chiprop}
\end{equation}
understood as an analytically continued distribution, with local contact terms handled by local counterterms. For the spatial Fourier convention $\hat f(k)=\int dx\,e^{-ikx}f(x)$, its nonanalytic part is $N_\chi A_s|k|^s$, where~\cite{BRRZ2}
\begin{equation}
A_s=2\Gamma(-s)\cos\frac{\pi s}{2}=-c_s<0\qquad(0<s<2).
\label{eq:chiFourier}
\end{equation}
The value at $s=1$ is understood by continuity. The corresponding homogeneous inverse kernel is formally
\begin{equation}
S_\chi=\frac{1}{2N_\chi A_s}\int d\tau\int\frac{dk}{2\pi}\,\chi(\tau,k)|k|^{-s}\chi(\tau,-k).
\label{eq:Schi}
\end{equation}
Its negative coefficient reflects the Fourier transform of the continued distribution. Equation~\eqref{eq:Schi} is a formal inverse kernel, not a positive Gaussian measure over real fields by itself; the perturbative definition is Wick contraction with Eq.~\eqref{eq:chiprop} and the stated local subtractions.
This $\chi$ is a GFF of the one-dimensional conformal group acting on each time slice, with dimension $(1+s)/2$ on the slice and no correlation between slices; it is Gaussian, but it is not a GFF of the two-dimensional conformal group. Coupling it locally to the Ising spin,
\begin{equation}
S=S_{\rm Ising_2}+S_\chi-g\int d^2X\,\sigma(X)\chi(X),\qquad X=(\tau,x),
\label{eq:S}
\end{equation}
and performing the Gaussian contractions reproduces \eqref{eq:SLR} with $\lambda=g^2N_\chi$. We choose the interaction sign $-g\sigma\chi$, so the Euclidean expansion contains $\exp(+g\int\sigma\chi)$; the transformation $\chi\to-\chi$ reverses this convention without changing the physics. A $\chi$ with a local $(\partial_\tau\chi)^2$ term would instead generate a retarded interaction that the lattice model does not have, so \eqref{eq:chiprop} is forced. At the decoupled point $z=1$, the correlator \eqref{eq:chiprop} scales as $b^{-(2+s)}$, so $\Delta^{(0)}_\chi=(2+s)/2$ and
\begin{equation}
\Delta_{\mathcal O}=\tfrac18+\tfrac{2+s}2=2-\tfrac\delta2,\qquad\mathcal O=\sigma\chi ,
\end{equation}
marginal at $s_*$ and weakly relevant below it, with the same $\mathbb Z_2^\sigma\times\mathbb Z_2^\chi$ structure as in the classical case:
\begin{equation}
\beta_g=-\frac\delta2 g+B_g g^3+O(g^5,\delta g^3).
\label{eq:betaform}
\end{equation}
The proposed quantum duality (Fig.~\ref{fig:duality}) identifies the fixed point of \eqref{eq:S} at $g_*^2=O(\delta)$ with the LRFP of \eqref{eq:Sphi}. The dictionary is the anisotropic version of the classical one. Nonrenormalization of the nonlocal kernel gives $\Delta_\chi=(1+z+s)/2$ exactly; the equation of motion gives $\Delta_\chi=\Delta_\sigma+s$; hence
\begin{equation}
\Delta_\sigma=\frac{1+z-s}2,\quad\Delta_\chi=\frac{1+z+s}2,\quad\Delta_\sigma+\Delta_\chi=1+z ,
\label{eq:shadowfull}
\end{equation}
which is Eq.~\eqref{eq:shadow} together with its shadow partner. The order parameter and thermal scaling operator are identified as $\phi\leftrightarrow\sigma$ and $[\phi^2]\leftrightarrow\varepsilon$. Extending the dictionary to the cubic composite requires more care. At criticality the bare Lifshitz equation of motion contains temporal and spatial derivatives,
\begin{equation}
\frac{u}{6}\phi^3=\partial_\tau^2\phi-c_s(-\partial_x^2)^{s/2}\phi+c_2\partial_x^2\phi.
\label{eq:quantumEOM}
\end{equation}
The temporal descendant and the fractional spatial derivative have dimensions $\Delta_\phi+2z$ and $\Delta_\phi+s$, respectively, which generally differ at the interacting fixed point. An identification of a renormalized cubic operator with $\chi$ therefore requires an operator-mixing analysis. We leave this extension open; the perturbative results below use the auxiliary-field shadow relation \eqref{eq:shadowfull}.

Two features distinguish the quantum construction from the classical one, and both follow from the $\delta(\tau)$ in \eqref{eq:chiprop}. The first concerns symmetry. The decoupled theory Ising$_2\otimes\chi$ is invariant under Euclidean rotations only in its Ising factor; the coupling $g$ transmits the breaking to the Ising sector, and the velocity of the Ising CFT---an exactly marginal, redundant modulus at $g=0$, corresponding to the spin-two component $T+\bar T$ of the stress tensor---becomes a running coupling. Its flow at the fixed point is what makes $z\neq1$: the dynamical exponent is the anomalous dimension of time. Recombination of the stress tensor therefore has a new component here: not only does $\partial^\mu T_{\mu\nu}$ fail to vanish, as in the classical case, but $T+\bar T$ itself appears in the $\mathcal O\times\mathcal O$ OPE with a nonzero integrated coefficient, which rotation invariance forbids classically. The second feature concerns the integrals. Every Wick contraction of two $\chi$'s forces the two spins they are attached to onto a common time slice, so the correlators entering the beta function are integrated over equal-time configurations only, with a purely spatial kernel. Table~\ref{tab:dict} summarizes the correspondence.

\begin{table}[t]
\caption{The classical and quantum crossover dualities side by side. In both cases $\beta_g=-\frac\delta2g+Bg^3$ with $\delta=s_*-s$, $N_\chi=1$, and $\sigma$ normalized to $\langle\sigma\sigma\rangle=|X|^{-2\Delta_\sigma}$. Classical entries are from Refs.~\cite{BRRZ1,BRRZ2}; $I_q=-0.4037456(1)$ is the shape integral \eqref{eq:Iq}.}
\label{tab:dict}
\setlength{\tabcolsep}{2pt}
\begin{ruledtabular}
\begin{tabular}{lcc}
 & classical, $d$ dim. & quantum chain\\
\hline
$\langle\chi\chi\rangle_0$ & $|x|^{-d-s}$ & $\delta(\tau)\,|x|^{-1-s}$\\
LRFP symmetry & conformal, no local $T$ & scale, $z\neq1$\\
crossover $s_*$ & $2-\eta_{\rm SR}$ & $2-\eta_{\rm SR}=7/4$\\
$\Delta_\chi$ (exact) & $(d+s)/2$ & $(1+z+s)/2$\\
$\Delta_\sigma$ (exact) & $(d-s)/2$ & $(1+z-s)/2$\\
$\Delta_\sigma+\Delta_\chi$ & $d$ & $1+z$\\
counterterms & $1,\varepsilon,\sigma\chi$ & $1,\varepsilon,\sigma\chi,T\!+\!\bar T$\\
$B$, $d=2$ & $-\pi I_q$ & $-I_q+7\pi/32$\\
 & $=1.268404$ & $=1.090969$\\
$\gamma^{(x)}_\varepsilon$, $O(g^2)$, $d=2$ & $0$ & $-\pi g^2/4$\\
$\Delta_\sigma-\tfrac18$, $O(\delta)$, $d=2$ & $\delta/2$ & $0.140\,\delta$\\
$1/\nu-1$, $O(\delta)$, $d=2$ & $0$ & $-0.360\,\delta$\\
$\omega$ & $\delta$ & $\delta$\\
$z-1$ & -- & $-0.720\,\delta$\\
\end{tabular}
\end{ruledtabular}
\end{table}

\subsection{Perturbation theory around the crossover}
\label{sec:pt}

We now compute the leading nontrivial terms of the RG flow of \eqref{eq:S} in conformal perturbation theory~\cite{Zamolodchikov1987,Cardy1996}. The input is exact: the Ising OPE coefficients and four-point functions~\cite{BPZ1984}, and Wick's theorem for $\chi$. Through $O(g^3)$ the divergences are canceled by four counterterms: the identity (vacuum energy), $\varepsilon$ (the position of the critical surface), $\sigma\chi$ (the coupling itself), and $T+\bar T$ (the velocity), the last of which has no classical counterpart. We work at $\delta=0$ in the shape integrals and extract poles in $\delta$ from the overall scale integrals; at this order the scheme dependence is confined to the finite parts of the counterterms and does not affect $z$, $\nu$, or $\omega$ at $O(\delta)$.

\subsubsection{Cubic beta function from the integrated four-spin function}
\label{sec:beta}

Because $C_{\mathcal{OOO}}=0$, the quadratic term in the beta function is absent and the first nontrivial term is cubic. To determine the renormalization of $g$, we extract the coefficient of $\mathcal O$ in the short-distance expansion of the perturbing insertions~\cite{KomargodskiSD,BRRZ1}. This coefficient can be obtained by inserting $\mathcal O(\infty)$ in the perturbative expansion of $|0\rangle_{g,V}=\exp\big(g_B\int_V\mathcal O\big)|0\rangle$ for a large region $V$:
\begin{align}
\frac{\langle\mathcal O(\infty)|0\rangle_{g,V}}{V}\sim{}& g_B+\frac{g_B^2}{2}\!\int\!\langle\mathcal O(\infty)\mathcal O(0)\mathcal O(X_1)\rangle\nonumber\\
&+\frac{g_B^3}{3!}\!\int\!\langle\mathcal O(\infty)\mathcal O(0)\mathcal O(X_1)\mathcal O(X_2)\rangle+\dots
\label{eq:A}
\end{align}
Here $\mathcal O(\infty)$ denotes a time-integrated insertion at spatial infinity. Inside correlators of the undeformed theory, with all other insertions held fixed, we define
\[
\mathcal O(\infty)\equiv\lim_{R\to\infty}
\frac{R^{1+s+2\Delta_\sigma}}{N_\chi}
\int_{-\infty}^{\infty}d\tau\,\mathcal O(\tau,x=R).
\]
When the distant $\chi$ contracts with $\chi(\tau_i,x_i)$, the time integral removes the delta function in \eqref{eq:chiprop}. The factor $R^{1+s}/N_\chi$ cancels the remaining propagator in the limit, leaving the usual Ising insertion $\sigma(\infty)=\lim_{R\to\infty}R^{2\Delta_\sigma}\sigma(\tau_i,R)$. Thus $\langle\mathcal O(\infty)\mathcal O(0)\rangle_0=1$, giving the linear term in \eqref{eq:A}. The quadratic term vanishes because it contains three $\chi$ fields. Before fixing one insertion at the origin, the cubic term is
\begin{equation}
\frac{g_B^3}{3!}\int_Vd^2Y_1\,d^2Y_2\,d^2Y_3\,\langle\mathcal O(\infty)\mathcal O(Y_1)\mathcal O(Y_2)\mathcal O(Y_3)\rangle ,
\label{eq:third}
\end{equation}
which we reduce in three steps.

For the integral below we set $N_\chi=1$ and restore it in the beta function.

(i) \emph{Wick contraction.} The distant $\chi$ can contract with any of the three insertions. In each case only the propagator between the other two remains:
\begin{align}
&\langle\mathcal O(\infty)\mathcal O(Y_1)\mathcal O(Y_2)\mathcal O(Y_3)\rangle=\langle\sigma(\infty)\sigma(Y_1)\sigma(Y_2)\sigma(Y_3)\rangle\nonumber\\
&\quad\times\big[\langle\chi(Y_2)\chi(Y_3)\rangle+\langle\chi(Y_1)\chi(Y_3)\rangle+\langle\chi(Y_1)\chi(Y_2)\rangle\big].
\label{eq:wick3}
\end{align}

(ii) \emph{Relabeling.} The Ising four-point function and the integration domain are symmetric under permutations of $Y_1,Y_2,Y_3$, so the three pairings in \eqref{eq:wick3} give equal contributions after renaming the integration variables: the sum is three times the first term. Each contracted pair is placed at equal time by the $\delta(\tau)$ in \eqref{eq:chiprop} and weighted by $|y_2-y_3|^{-1-s}$.

(iii) \emph{Fixing the origin.} The integrand is translation invariant, so one integration only produces the volume $V$; we use it to put the insertion paired with the distant $\chi$ at the origin, $Y_1\to0$, $Y_2\to X_1$, $Y_3\to X_2$. Dividing by $V$ and using (ii), the third-order term of \eqref{eq:A} becomes
\begin{equation}
\frac{g_B^3}{2}\int d^2X_1\,d^2X_2\,\langle\sigma(\infty)\sigma(0)\sigma(X_1)\sigma(X_2)\rangle\,\frac{\delta(\tau_1-\tau_2)}{|x_1-x_2|^{1+s}} ,
\label{eq:Jq}
\end{equation}
where the $\delta$ removes one of the two time integrals. What remains is the integral over the positions of the pair and its common time $\tau$ relative to the operator at the origin,
\begin{equation}
J_q=\int d\tau\,dx\,dy\,
\frac{\langle\sigma(\infty)\sigma(0)\sigma(\tau,x)\sigma(\tau,y)\rangle}{|x-y|^{1+s}} .
\end{equation}
The third-order term equals $g_B^3 J_q/2$ with both signs of $r=x-y$ included, or equivalently $g_B^3 J_q|_{r>0}$.

With $\langle\sigma(X)\sigma(0)\rangle=|X|^{-1/4}$, introduce the signed separation $r=x-y$ and the center $c=(x+y)/2$ of the equal-time pair, and the dimensionless shape variable $w=p+iq$, $p=c/r$, $q=\tau/r$. Then $d\tau\,dx\,dy=|r|^2d^2w\,dr$, conformal covariance supplies $|r|^{-1/4}$, and
\begin{equation}
|r|^2\cdot |r|^{-1-s}\cdot |r|^{-1/4}\,dr=|r|^{-1+\delta}\,dr .
\end{equation}
The scale integral $\int_0^{\mu^{-1}}dr\,r^{-1+\delta}=\delta^{-1}-\ln\mu+O(\delta)$ carries the pole---the simultaneous collision of the three perturbing operators---and its residue is a pure number, the integral over the shape plane of
\begin{align}
F_q(w)&=\langle\sigma(\infty)\sigma(0)\sigma(w+\tfrac12)\sigma(w-\tfrac12)\rangle\nonumber\\
&=\frac{\sqrt{(1+r_++r_-)/2}}{(r_+r_-)^{1/4}},\qquad r_\pm=|w\pm\tfrac12|.
\label{eq:ising4}
\end{align}
The collisions of either spin with the origin, at $w=\pm\tfrac12$ are integrable. The large-$|w|$ behavior is governed by the $\sigma\times\sigma$ OPE,
\begin{equation}
F_q=1+\frac{1}{4R}+\frac{\cos 2\theta}{32R^2}+O(R^{-3}),\quad w=Re^{i\theta},
\label{eq:Gope}
\end{equation}
whose first two terms give the two power-law subdivergences associated with the pair collision. The identity gives $\pi R^2$ and renormalizes the vacuum energy; the $\varepsilon$ term ($C_{\sigma\sigma\varepsilon}^2=\tfrac14$) gives $\pi R/2$ and shifts the critical surface. Neither contributes to the beta function for $g$. The stress-tensor term of order $R^{-2}$ would give a logarithm, but its angular average vanishes. After these subtractions, the coefficient of the overall logarithm is determined by the finite shape integral
\begin{align}
I_q&=\lim_{R\to\infty}\left[\int_{|w|<R}d^2w\,F_q(w)-\pi R^2-\frac\pi2 R\right]\nonumber\\
&=-0.4037456(1),
\label{eq:Iq}
\end{align}
evaluated numerically using Gauss--Legendre quadrature~\cite{DLMFQuadrature} on subdomains chosen so that the two singularities lie on their boundaries, and independently checked by adaptive numerical quadrature (see Ref.~\cite{QUADPACK} for adaptive integration methods). Irrelevant operators such as $\chi^2$ and $\varepsilon\chi^2$ have dimension above two at the decoupled point and need no counterterms. Restoring $N_\chi$, the pole is canceled by $g_B=\mu^{\delta/2}(g-N_\chi I_qg^3/\delta+\dots)$, so that
\begin{align}
\mu\frac{dg}{d\mu}&=-\frac\delta2g+B_0g^3,\nonumber\\
B_0&=-N_\chi I_q=0.4037456(1)N_\chi.
\label{eq:B0}
\end{align}
This is the flow in coordinates in which the Ising velocity is allowed to run; it is not yet the physical cubic coefficient.

The relation to the classical $d=2$ calculation is now transparent. There the contracted pair is separated by a two-vector $r$ and the shape variable is the complex ratio $c/r$; by conformal covariance the four-spin function is again $|r|^{-1/4}F_q(w)$ with the same function \eqref{eq:ising4}, and the only difference is the scale integral, $\int d^2r\,|r|^{-2+\delta}$ with angular measure $2\pi$ instead of $\int dr\,|r|^{-1+\delta}$ with the two signs of $r$. The classical cubic coefficient is therefore
\begin{equation}
\beta_3=-\pi I_q=1.268404(1),
\end{equation}
in agreement with $1.2684040(5)$ of Ref.~\cite{BRRZ1}. This is a check of both computations, and it isolates what is new in the quantum problem: not the shape integral, which happens to coincide in $d=2$, but the measure and, above all, the anisotropy to which we now turn.

\subsubsection{Anisotropy and the dynamical exponent}
\label{sec:aniso}

In Sec.~\ref{sec:beta}, the coefficient of $\mathcal O$ in the short-distance expansion of three perturbing insertions determined the flow of $g$. The velocity flow instead follows from the coefficient of $T$ in the OPE of two perturbing operators, which we extract using an insertion of $T(\infty)$. Reflection symmetry gives the same coefficient for $\bar T$, so the induced deformation is proportional to $T+\bar T$. With the velocity measured in units of the undeformed Ising velocity, an infinitesimal change contributes $-(\ln v_B)/(2\pi)\int(T+\bar T)$ to the action. To the order needed, the deformed vacuum is
\begin{equation}
|0\rangle_{B,V}=\exp\Big[\frac{\ln v_B}{2\pi}\int_V(T+\bar T)+g_B\int_V\mathcal O\Big]|0\rangle .
\end{equation}
Its overlap with the state created by $T$ is the analogue of Eq.~\eqref{eq:A}:
\begin{align}
\frac{\langle T(\infty)|0\rangle_{B,V}}{V}\sim{}&\frac{c}{4\pi}\ln v_B\nonumber\\
&+\frac{g_B^2}{2}\!\int\! d^2X_1\,\langle T(\infty)\mathcal O(0)\mathcal O(X_1)\rangle+\dots
\label{eq:AT}
\end{align}
The first term is the tree-level overlap of the counterterm with its own state,
\begin{equation}
\frac{\ln v_B}{2\pi}\int_Vd^2X\,\langle T(\infty)(T+\bar T)(X)\rangle=\frac{c}{4\pi}\ln v_B\,V,
\end{equation}
since $T(\infty)=\lim_{z\to\infty}z^4T(z)$, $\langle T(\infty)T(X)\rangle=c/2$, and $\langle T(\infty)\bar T(X)\rangle=0$. Thus $c/(4\pi)$ plays the role of the unit overlap in Eq.~\eqref{eq:A}. The term linear in $g_B$ vanishes because it contains a single $\chi$.

Before fixing one insertion at the origin in Eq.~\eqref{eq:AT}, introduce the pair center and signed spatial separation $r=x_1-x_2$, as in Sec.~\ref{sec:beta}. Translation invariance gives the factor $V$ from the center integral. The only Wick contraction is between the two $\chi$ fields: their propagator fixes the relative time to zero and leaves $N_\chi |r|^{-1-s}$. The remaining Ising correlator is fixed by the Ward identity,
\begin{equation}
\langle T(\infty)\sigma(0)\sigma(z)\rangle=h_\sigma\,z^2\,|z|^{-1/4},\qquad h_\sigma=\tfrac1{16} ,
\label{eq:ward}
\end{equation}
equivalently the stress-tensor term of the Ising OPE,
\begin{equation}
\sigma(z)\sigma(0)=|z|^{-1/4}\Big[1+\tfrac12|z|\,\varepsilon+\tfrac{2h_\sigma}{c}\big(z^2T+\bar z^2\bar T\big)+\dots\Big],
\label{eq:OPE}
\end{equation}
with $c=1/2$ and $2h_\sigma/c=1/4$. On the spatial axis $z=r$, so the three-point function is $h_\sigma |r|^{7/4}$. There is now only one separation and hence no pair subdivergence or continuous shape variable. The two signs of $r$ contribute equally and cancel the factor $1/2!$ in Eq.~\eqref{eq:AT}. Writing $\rho=|r|$, the quadratic term in Eq.~\eqref{eq:AT} becomes
\begin{align}
\frac{g_B^2}{2}\!\int\! d^2X_1&\langle T(\infty)\mathcal O(0)\mathcal O(X_1)\rangle\nonumber\\
&=\frac{g_B^2N_\chi h_\sigma}{2}\sum_{\pm}\int_a^{\mu^{-1}}\! d\rho\,\rho^{-1+\delta}\nonumber\\ &\xrightarrow{\delta\to0}g_B^2N_\chi h_\sigma\ln\frac{1}{a\mu} ,
\label{eq:ATlog}
\end{align}
Equivalently, analytic regularization gives $g_B^2N_\chi h_\sigma\mu^{-\delta}/\delta$, with the same upper limit as in Sec.~\ref{sec:beta}. Hence, using $4\pi h_\sigma/c=\pi/2$,
\begin{equation}
\frac{\langle T(\infty)|0\rangle_{B}}{V}\sim\frac{c}{4\pi}\Big[\ln v_B+\frac{\pi N_\chi}{2}\,\frac{g_B^2}{\delta}\,\mu^{-\delta}\Big].
\label{eq:ATpole}
\end{equation}

The above calculation differs from its classical long-range Ising model counterpart in the angular integral. For a rotation-invariant kernel one must integrate over the orientation $\varphi$ of $z=\rho e^{i\varphi}$. The radial integral is the same, but the spin-two phase in Eq.~\eqref{eq:ward} gives
\begin{align}
\int d^2z\,&|z|^{-2-s}\langle T(\infty)\sigma(0)\sigma(z)\rangle\nonumber\\
&=h_\sigma\int d\rho\,\rho^{-1+\delta}\int_0^{2\pi}\!d\varphi\,e^{2i\varphi}=0 .
\label{eq:ATcl}
\end{align}
For the equal-time kernel the only orientations are $0$ and $\pi$, and both give $e^{2i\varphi}=1$. This is also why the angular cancellation in Eq.~\eqref{eq:Gope} does not remove the velocity pole: there the angle is that of the pair center relative to a third insertion; here there is no third insertion.

To interpret the residue of Eq.~\eqref{eq:ATpole}, consider a scalar Ising primary $\Phi$ at separation $X$, making an angle $\theta$ with the spatial axis. The integrated stress-tensor Ward identity gives $\delta\ln\langle\Phi(X)\Phi(0)\rangle=\Delta_\Phi\cos(2\theta)\,\delta\ln v$ for the action variation above. This is the response to the area-preserving change of coordinates $x'=v^{-1/2}x$, $\tau'=v^{1/2}\tau$. In these coordinates the Ising correlator is the unit-velocity one:
\begin{align}
\langle\Phi(x,\tau)\Phi(0)\rangle&=\big(v^{-1}x^2+v\,\tau^2\big)^{-\Delta_\Phi}\nonumber\\
&=v^{\Delta_\Phi}\big(x^2+v^2\tau^2\big)^{-\Delta_\Phi};
\label{eq:squeezedcorr}
\end{align}
in particular its equal-time normalization is $v^{\Delta_\Phi}$. At $g=0$ the squeeze is just a choice of coordinates. At nonzero $g$ the instantaneous kernel singles out the spatial axis, and the velocity counterterm is fixed by the pole. Finiteness of Eq.~\eqref{eq:ATpole} requires $\ln v_B=\ln v(\mu)-\pi N_\chi g^2/(2\delta)+O(g^4)$. Differentiating at fixed bare couplings and using $\beta_g=-\delta g/2+O(g^3)$ gives
\begin{equation}
\mu\frac{d\ln v}{d\mu}=C_ag^2+O(g^4),\qquad C_a=-\frac{N_\chi\pi}{2} .
\label{eq:Ca}
\end{equation}
In spatial RG variables the characteristic frequency scales as $v(k)|k|\propto |k|^z$. Converting the isotropic subtraction scale used above to a spatial momentum scale changes the velocity beta function first at $O(g^4)$. Thus, to the order computed here,
\begin{equation}
z=1+\beta_{\ln v}\big|_*+O(\bar g_*^4).
\label{eq:zfromv}
\end{equation}

The velocity counterterm couples to $T+\bar T$ of the Ising sector. Before we change coordinates, the spin correlator has therefore already responded to this deformation through the Ising Ward identity. At equal time, Eq.~\eqref{eq:squeezedcorr} gives $v^{\Delta_\sigma^{(0)}}|x|^{-2\Delta_\sigma^{(0)}}$. Setting $x=\sqrt v\,x'$ absorbs precisely this response into the separation, leaving $|x'|^{-2\Delta_\sigma^{(0)}}$ with unit coefficient.

The auxiliary field $\chi$ belongs to the other factor of the decoupled theory. An insertion of the Ising stress tensor in its normalized two-point function factorizes and gives zero connected response. Its prescribed covariance therefore still has the form in Eq.~\eqref{eq:chiprop} in the original coordinates. Expressing that kernel in the same primed coordinates gives
\begin{equation*}
N_\chi\frac{\delta(\tau)}{|x|^{1+s}}
= N_\chi v^{-s/2}\frac{\delta(\tau')}{|x'|^{1+s}} .
\end{equation*}
The time delta function contributes $v^{1/2}$ and the spatial power contributes $v^{-(1+s)/2}$. Restoring the prescribed coefficient $N_\chi$ in the primed coordinates therefore requires $\chi'(X')=v^{s/4}\chi(X)$. The normalized operator is then $\mathcal O'(X')=\sigma'(X')\chi'(X')=v^{s/4}\mathcal O(X)$. Since $d^2X'=d^2X$, the coupling multiplying this operator is fixed by the interaction,
\begin{equation*}
g\int d^2X\,\mathcal O(X)=\bar g\int d^2X'\,\mathcal O'(X'),
\qquad \bar g=g v^{-s/4}.
\end{equation*}
The running velocity enters the coupling flow through $\beta_{\bar g}/\bar g=\beta_g/g-(s/4)\beta_{\ln v}$, where $\beta_{\ln v}=\mu\,d\ln v/d\mu$. Using Eqs.~\eqref{eq:B0} and~\eqref{eq:Ca} and evaluating the cubic coefficient at $s_*=7/4$ gives
\begin{align}
\beta_{\bar g}&=-\frac\delta2\bar g+B_g\bar g^3,\nonumber\\
B_g&=B_0-\frac{7}{16}C_a=N_\chi\Big(-I_q+\frac{7\pi}{32}\Big)=1.090969(1)\,N_\chi .
\label{eq:Bg}
\end{align}
The fixed point is at $\bar g_*^2=\delta/2B_g=0.458308(1)\,\delta/N_\chi$, and the compensating squeeze that keeps $v=1$ is the anisotropic scaling of the fixed point. Equation \eqref{eq:zfromv} gives
\begin{align}
z&=1+C_a\bar g_*^2=1+\frac{C_a}{2B_g}\,\delta\nonumber\\
&=1-0.719909(1)\Big(\frac74-s\Big)+O(\delta^2).
\label{eq:z}
\end{align}
Both $C_a$ and $B_g$ are proportional to $N_\chi$, so the slope is normalization independent. The calculation predicts that $z$ approaches $1$ from below; continuity alone, as used in Sec.~\ref{sec:sak}, fixes the endpoint value.

\subsubsection{Scaling dimension of the thermal operator}
\label{sec:thermal}

The equal-time correlator $\langle\varepsilon(x,0)\varepsilon(0,0)\rangle\propto |x|^{-2\Delta_\varepsilon}$ defines the spatial thermal dimension. Analogously to Eq.~\eqref{eq:A}, we expand the normalized two-point function $G_\varepsilon(X)\equiv\langle\varepsilon(X)\varepsilon(0)\rangle_g$ in the perturbation $\mathcal O=\sigma\chi$. The linear term vanishes because it contains a single $\chi$, and the regulated expansion is
\begin{align}
G_\varepsilon(X)={}&G_\varepsilon^{(0)}(X)
+\frac{g_B^2}{2}\int d^2Y_1\,d^2Y_2\nonumber\\
&\times\langle\chi(Y_1)\chi(Y_2)\rangle_0\nonumber\\
&\times\langle\varepsilon(X)\varepsilon(0)\sigma(Y_1)\sigma(Y_2)\rangle_{0,\mathrm{sub}}
+O(g_B^4).
\label{eq:thermalexpansion}
\end{align}
Here $G_\varepsilon^{(0)}(X)=|X|^{-2}$. Throughout this and the next subsection, $\mathrm{sub}$ denotes the subtraction required by the partition-function normalization:
\begin{align}
&\langle\Phi(X)\Phi(0)\sigma(Y_1)\sigma(Y_2)\rangle_{0,\mathrm{sub}}\nonumber\\
&\quad=\langle\Phi(X)\Phi(0)\sigma(Y_1)\sigma(Y_2)\rangle_0\nonumber\\
&\qquad-G_\Phi^{(0)}(X)\langle\sigma(Y_1)\sigma(Y_2)\rangle_0,
\label{eq:vacuumsubtraction}
\end{align}
where $G_\Phi^{(0)}(X)=\langle\Phi(X)\Phi(0)\rangle_0$ and $\Phi=\varepsilon$ or $\sigma$ labels the external operator. Local counterterms are understood in both expansions. The $\chi$ contraction in Eq.~\eqref{eq:chiprop} sets the two perturbing spins at equal time. They can approach either energy insertion, and the coefficient of the resulting logarithm at each insertion is proportional to the shape integral $I_\varepsilon$. Set the spin separation to one and denote the pair center by $w$. The mixed Ising four-point function then gives~\cite{BRRZ1}
\[
I_\varepsilon=\lim_{R\to\infty}\int_{|w|<R}d^2w
\left(\frac{|w|^2}{|w^2-1/4|}-1\right),
\]
where the $-1$ implements Eq.~\eqref{eq:vacuumsubtraction} in these coordinates. In elliptic coordinates $w=\tfrac12\cosh(\eta+i\theta)$, $\eta\geq0$, $0\leq\theta<2\pi$ the Jacobian is $|w^2-1/4|$, so the subtracted integrand, including the measure, reduces to $\tfrac14\cos(2\theta)\,d\eta\,d\theta$. Its angular average vanishes on every ellipse. The circular cutoff gives the same limit: at large $R$ the difference in area between the disk and an ellipse with semimajor axis $R$ is bounded, while the subtracted integrand is $O(R^{-2})$ there. Thus $I_\varepsilon=0$.

The remaining correction comes from the running velocity. Since $\Delta_\varepsilon^{(0)}=1$, Eq.~\eqref{eq:squeezedcorr} gives an equal-time correlator proportional to $v/|x|^2$. At the fixed point $v(\mu)\propto\mu^{z-1}$ to this order by Eq.~\eqref{eq:zfromv}. Setting $\mu=1/|x|$ and reading off the power of $|x|$ gives
\begin{align}
\Delta_\varepsilon
&=1+\tfrac12(z-1)+O(\delta^2)\nonumber\\
&=1-0.359954(1)\,\delta+O(\delta^2).
\label{eq:Deps}
\end{align}
The coupling to $\varepsilon$ has scaling dimension $1+z-\Delta_\varepsilon$, hence
\begin{equation}
\frac1\nu=1+z-\Delta_\varepsilon=\frac{1+z}{2}+O(\delta^2).
\label{eq:nu}
\end{equation}

\subsubsection{Spin dimension and correlator checks}
\label{sec:checks}

For the spin correlator $G_\sigma(X)\equiv\langle\sigma(X)\sigma(0)\rangle_g$, the corresponding expansion is
\begin{align}
G_\sigma(X)={}&G_\sigma^{(0)}(X)
+\frac{g_B^2}{2}\int d^2Y_1\,d^2Y_2\nonumber\\
&\times\langle\chi(Y_1)\chi(Y_2)\rangle_0\nonumber\\
&\times\langle\sigma(X)\sigma(0)\sigma(Y_1)\sigma(Y_2)\rangle_{0,\mathrm{sub}}
+O(g_B^4).
\label{eq:spinexpansion}
\end{align}
Here $G_\sigma^{(0)}(X)=|X|^{-1/4}$, and the subtraction is defined in Eq.~\eqref{eq:vacuumsubtraction}. It retains the two crossed products pairing each external spin with an integrated spin, so $\mathrm{sub}$ is not the fully connected four-point cumulant. For the thermal correlator these crossed products vanish because the mixed two-point function $\langle\varepsilon\sigma\rangle_0$ vanishes.

We now use $G_\Phi(X)\equiv\langle\Phi(X)\Phi(0)\rangle_g$ to discuss both cases together. The external operators $\Phi=\sigma,\varepsilon$ are scalar primaries of the undeformed Ising CFT: both have zero conformal spin, $h_\Phi-\bar h_\Phi=0$. Their internal $\mathbb Z_2$ parity differs: the spin primary $\sigma$ is odd and has $\Delta_\sigma^{(0)}=1/8$, while the energy primary $\varepsilon$ is even and has $\Delta_\varepsilon^{(0)}=1$. The two integrated insertions are always $\sigma(Y_1)\sigma(Y_2)$, as fixed by the perturbation $\mathcal O=\sigma\chi$.

The angular dependence of the resulting correlator also provides an independent check of the velocity flow. At $O(g^2)$ there are two logarithmic regions. When the perturbing pair collapses onto an external insertion, each end of the correlator contributes the shape integral $I_\Phi$, with $I_\sigma=I_q$ and $I_\varepsilon=0$ as computed above. When the pair collapses away from both insertions, its OPE generates $T+\bar T$. The Ward identity used in deriving Eq.~\eqref{eq:squeezedcorr} gives this contribution an angular dependence $\cos 2\vartheta$, where $\vartheta$ is the angle between the external separation $X$ and the spatial axis. Adding the two regions gives, with $\rho=|X|$,
\begin{equation}
\frac{G_\Phi(X)}{G_\Phi^{(0)}(X)}=1+N_\chi g^2\Big[2I_\Phi+\frac{\pi\Delta^{(0)}_\Phi}{2}\cos2\vartheta\Big]\ln\rho+\dots
\label{eq:angularlog}
\end{equation}
This is precisely the expansion of the anisotropic scaling form $x^{-2\Delta_\Phi}f(\tau/x^z)$ with $z=1+C_ag^2$ and
\begin{equation}
\gamma^{(x)}_\Phi\equiv\Delta_\Phi-\Delta_\Phi^{(0)}=-N_\chi I_\Phi g^2+\frac{\Delta_\Phi^{(0)}}{2}C_ag^2+O(g^4).
\label{eq:gammaphys}
\end{equation}
The angular coefficient fixes $C_a=-\pi N_\chi/2$ independently of the velocity calculation. To check Eq.~\eqref{eq:angularlog} numerically, we place the external operators at $0,1$ on the spatial axis and define, for $r>0$,
\begin{equation}
Q_\Phi(c,r)=r^{1/4}\langle\Phi(0)\Phi(1)\sigma(c+\tfrac r2)\sigma(c-\tfrac r2)\rangle_0.
\label{eq:finitePair}
\end{equation}
The vacuum subtraction removes $1$. For external spins the pair OPE also gives $r/[4|c(1-c)|]$ at fixed $c\neq0,1$. We subtract this thermal contribution using the same critical-surface prescription as in Sec.~\ref{sec:beta}. The integrals used in the check are therefore
\begin{align}
J_\sigma(r)&=\frac{1}{r^2}\int d^2c\left[Q_\sigma(c,r)-1-\frac{r}{4|c(1-c)|}\right],\nonumber\\
J_\varepsilon(r)&=\frac{1}{r^2}\int d^2c\left[Q_\varepsilon(c,r)-1\right].
\label{eq:finitePairSubtractions}
\end{align}
The thermal correlator has no corresponding energy-channel term because $C_{\varepsilon\varepsilon\varepsilon}=0$. Integrating the subtracted expressions at finite $r$ and extrapolating $r\to0$ gives $J_\sigma\to2I_q+\pi/16=-0.61114$ and $J_\varepsilon\to\pi/2$, reproducing the logarithmic coefficients for spatial separation in Eq.~\eqref{eq:angularlog}. These limits include both the endpoint and stress-tensor regions.

Specializing to $\Phi=\sigma$, with $\Delta_\Phi^{(0)}=1/8$ and $I_\Phi=I_q$, gives
\begin{align}
\Delta_\sigma&=\frac18+N_\chi\Big(-I_q-\frac{\pi}{32}\Big)\bar g_*^2+O(\delta^2)\nonumber\\
&=\frac18+0.140046(1)\,\delta+O(\delta^2),
\label{eq:Dsigma}
\end{align}
Together with Eq.~\eqref{eq:z}, this satisfies the exact shadow relation \eqref{eq:shadowfull}, $2\Delta_\sigma=1+z-s$. At this order the equality reduces to $-2N_\chi I_q-7C_a/8=2B_g$, which follows directly from Eq.~\eqref{eq:Bg}. This tests the consistency of the coupling and operator renormalizations; the angular term in Eq.~\eqref{eq:angularlog} separately checks the velocity residue.

\subsubsection{Correction to scaling}
\label{sec:corrections}

Linearizing \eqref{eq:Bg} at the fixed point,
\begin{equation}
\omega=\frac{d\beta_{\bar g}}{d\bar g}\Big|_{\bar g_*}=\delta+O(\delta^2),
\label{eq:omega}
\end{equation}
as in the classical case. The leading irrelevant operator at the LRFP is the $\mathbb Z_2$-even scalar conjugate to the flowing coupling---$\sigma\chi$ near $s_*$, $\phi^4$ near $2/3$---and its dimension $1+z+\omega$ approaches the measure dimension at both ends, with $\omega\simeq\epsilon$ near the mean-field endpoint and $\omega\simeq\delta$ near the short-range crossover. Finite-size corrections therefore decay as $L^{-\delta}$ near the crossover, which is slow: at $s=1.5$, $L^{-1/4}$ is still $0.35$ at $L=64$. At $s=s_*$ the coupling is marginally irrelevant and runs logarithmically. A free power-law fit to finite-size data near $s_*$ tends to round the endpoint cusp and bias $z$ downward, the pattern seen in the QMC data of Ref.~\cite{Shiratani2024}; Eq.~\eqref{eq:omega} is the input we use to control the extrapolation in Sec.~\ref{sec:dmrg}.

\subsubsection{Endpoint interpolation}
\label{sec:omega}

The Gaussian value $z=s/2$ is exact at $s=2/3$ but would give $z=7/8$ at the crossover, i.e.\ a discontinuous $z$: the interactions must lift $z$ above the Gaussian line by exactly $1/8$ at $s_*$, and the question is how $z-s/2$ grows in between. The two endpoint expansions fix its behavior at both ends. With $x=(s-\tfrac23)/\tfrac{13}{12}\in[0,1]$, Eq.~\eqref{eq:zMF} gives $z-s/2=Ax^2+O(x^3)$ with $A=\tfrac{169}{64}(0.0311)=0.08212$ as $x\to0$, and Eq.~\eqref{eq:z} gives $z-s/2=\tfrac18-K(1-x)+O((1-x)^2)$ with $K=\tfrac{13}{12}(0.719909-\tfrac12)=0.23823$ as $x\to1$. The simplest pole-free Pad\'e form that reproduces the vanishing value and slope at the Gaussian end and matches the remaining three numbers is
\begin{equation}
z-\frac s2=\frac{A\,x^2}{1+cx+dx^2},\qquad c=-0.7479,\quad d=0.4049,
\label{eq:pade}
\end{equation}
whose denominator has no real zeros. It gives $z=0.510$, $0.658$, and $0.823$ at $s=1$, $1.25$, and $1.5$. 
We emphasize that any two-point Pad\'e is an interpolation rather than a controlled prediction in the middle of the interval; taking other rational or polynomial leads to different predictions at the middle points. Testing it is the purpose of the next section.

\section{DMRG simulation}
\label{sec:dmrg}

\subsection{Lattice model, periodic images, and method}
\label{sec:method}

We simulate Eq.~\eqref{eq:H} with $J=1$ on periodic chains of $L$ sites, replacing the coupling by the sum over all periodic images,
\begin{align}
J_L(r)&=\sum_{n\in\mathbb Z}\frac1{|r+nL|^{1+s}}\nonumber\\
&=L^{-1-s}\Big[\zeta\Big(1+s,\frac rL\Big)+\zeta\Big(1+s,1-\frac rL\Big)\Big],
\label{eq:images}
\end{align}
with $\zeta$ the Hurwitz zeta function, so that no truncation of the interaction range is involved. This is the same Hamiltonian, with the same normalization, as in the QMC study of Ref.~\cite{Shiratani2024}, which allows a direct comparison of critical fields.

The ground state is obtained with the two-site DMRG algorithm as implemented in TeNPy~\cite{TeNPy}. The matrix-product operator (MPO) contains all $L-1$ distances explicitly, with bond dimension $L+2$.
The periodic Hamiltonian is represented by an open-boundary matrix-product state, which costs a doubling of the entanglement across the seam, $S\simeq\frac c3\ln L$ instead of $\frac c6\ln L$. 
Bond dimensions $\chi=64$ ($L\le32$), $96$ ($L=48$--$96$), and $128$ ($L=128$) give truncation errors below $2\times10^{-9}$. 
Sweeps are continued until the energy changes by less than $10^{-9}$ relative per sweep.
We conserve the Ising $\mathbb Z_2$ parity, $P=\prod_iX_i$, so that the finite-size gap $\Delta_L=E_0^{\rm odd}-E_0^{\rm even}$ is a difference of two ground-state energies obtained in separate symmetry sectors; no excited-state targeting is required. 
The implementation was validated against exact diagonalization at $L=10$, where energies in both sectors and all moments of the magnetization agree to ten digits. We study $L=16$, $24$, $32$, $48$, $64$, $96$, and $128$, i.e.\ five pairs $(L,2L)$, at $s=1$, $1.25$, and $1.5$.

\subsection{Observables}
\label{sec:obs}

In the even ground state we compute the full equal-time correlation matrix $C_{ij}=\langle Z_iZ_j\rangle$ (translation averaged, $C_L(r)$), from which follow the magnetization moment $m^2_L=L^{-2}\sum_{ij}C_{ij}$, the structure factor $S_L(k)=L^{-1}\sum_{ij}e^{ik(i-j)}C_{ij}$, and two dimensionless ratios: the structure-factor ratio $R_S=S_L(2\pi/L)/S_L(0)$ and the correlation ratio $R_C=C_L(L/2)/C_L(L/4)$, the spatial ratio used in Ref.~\cite{Shiratani2024}. 
For each $L$ we compute all observables on a grid of five to nine values of $\Gamma$ spanning $\pm5\%$ around the critical point, refined to a spacing of $0.01$--$0.02$ in the immediate vicinity of $\Gamma_c$ for the two or three largest sizes (the larger $L$, the narrower the window needed, because the crossings and the exponent estimates become increasingly sensitive to $\Gamma$), and we interpolate in $\Gamma$ with cubic splines. 

\begin{figure}[t]
\centering
\includegraphics[width=\columnwidth]{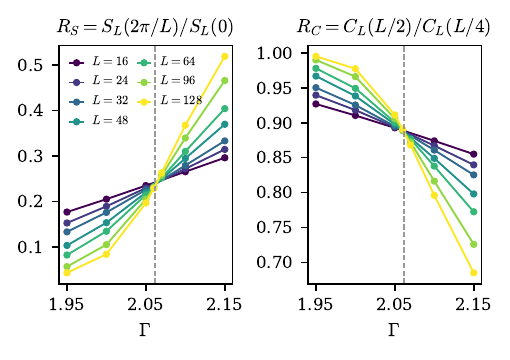}
\caption{Crossings of the two dimensionless ratios at $s=1.25$ for $L=16$--$128$. The dashed line marks the extrapolated $\Gamma_c=\GcTwo$. The structure-factor ratio approaches the crossing point from above, while the correlation ratio has essentially converged at $L=48$.}
\label{fig:crossings}
\end{figure}

\subsection{Locating the critical point}
\label{sec:Gc}

A dimensionless ratio $R_L(\Gamma)$ obeys $R_L=\mathcal R\big((\Gamma-\Gamma_c)L^{1/\nu}\big)+L^{-\omega}\mathcal R_1\big((\Gamma-\Gamma_c)L^{1/\nu}\big)+\dots$, and expanding both scaling functions to first order in $t=\Gamma-\Gamma_c$, the crossing of $R_L$ and $R_{2L}$ approaches $\Gamma_c$ as
\begin{equation}
\Gamma_\times(L,2L)=\Gamma_c+d_R\,L^{-(\omega+1/\nu)},\quad d_R=\frac{\mathcal R_1(0)}{\mathcal R'(0)}\,\frac{1-2^{-\omega}}{2^{1/\nu}-1},
\label{eq:crossdrift}
\end{equation}
the standard result for crossing-point sequences~\cite{Binder1981,Luck1985,Shao2016}: for a dimensionless quantity the value at $\Gamma_c$, $\mathcal R(0)$, is the same for every $L$, so the curves for $L$ and $2L$ intersect at $t\approx0$ and the crossing is displaced only by the correction term, by $L^{-\omega}$ divided by the slope $\propto L^{1/\nu}$.

\begin{figure*}[t]
\centering
\includegraphics[width=\textwidth]{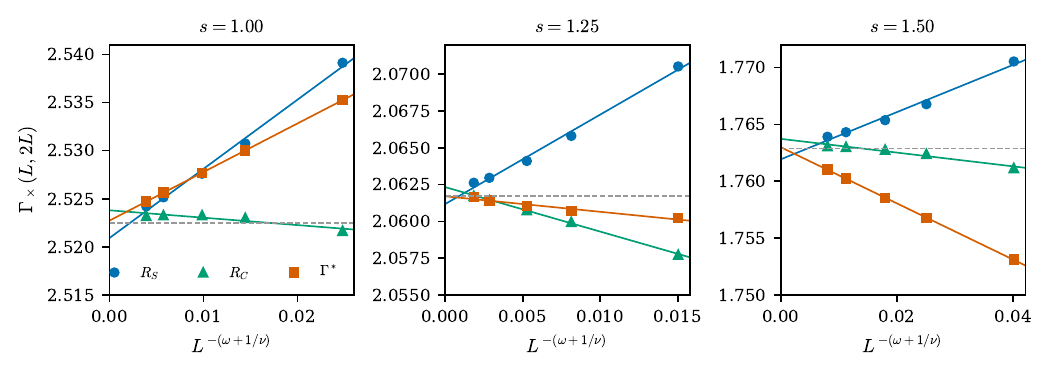}
\caption{Crossing points $\Gamma_\times(L,2L)$ of the structure-factor ratio $R_S$, the correlation ratio $R_C$, and the self-consistency point $\Gamma^*$ at which the gap and static estimates of $z$ coincide, plotted against $L^{-(\omega+1/\nu)}$ with $\omega$ and $1/\nu$ fixed to their theoretical inputs. Lines are the linear extrapolations of Eq.~\eqref{eq:crossdrift}; the dashed horizontal line is the adopted $\Gamma_c$, Eq.~\eqref{eq:Gcvalues}. The extrapolated values bracket $\Gamma_c$ from both sides.}
\label{fig:gcseq}
\end{figure*}

We use three such sequences (Figs.~\ref{fig:crossings} and \ref{fig:gcseq}, Table~\ref{tab:seq}): the crossings $\Gamma_\times(R_S)$ of the structure-factor ratio, which approach $\Gamma_c$ from above; the crossings $\Gamma_\times(R_C)$ of the correlation ratio, which increase toward their limit with a small amplitude and are nearly converged at $L=48$; and the self-consistency point $\Gamma^*(L,2L)$, the field at which the two finite-size estimates of $z$ of Sec.~\ref{sec:tworoutes}, Eqs.~\eqref{eq:zgap} and \eqref{eq:zD}, coincide. 
The last is a crossing in the same sense---the two estimators are dimensionless functions of the scaling variable with opposite responses to $\Gamma-\Gamma_c$---and obeys the same drift law; it approaches $\Gamma_c$ from below at $s=1.25$ and $1.5$ and from above at $s=1$. Each sequence is fitted to Eq.~\eqref{eq:crossdrift} with the exponents $\omega$ and $1/\nu$ fixed by the theory; only $\Gamma_c$ and $d_R$ are fitted. Near the mean-field endpoint on the interacting side, the expansion gives $\omega=\epsilon+O(\epsilon^2)$ and $1/\nu=s-\epsilon/3+O(\epsilon^2)$. 
For the finite-size fits we adopt the heuristic continuation $\omega_{\rm MF}=\min(\epsilon,\,2-s)$. The $2-s$ comes from the Gaussian kinetic term with $k^2$.
On the crossover side $\omega=\delta$ and $1/\nu=(1+z)/2=1-0.360\,\delta$ [Eqs.~\eqref{eq:omega} and \eqref{eq:nu}]. 
We use these mean-field-endpoint estimates at $s=1$ ($\omega=0.50$, $1/\nu=0.83$), the crossover values at $s=1.5$ ($\omega=0.25$, $1/\nu=0.91$), and at $s=1.25$, near the middle of the interval, the average of the two ($\omega=0.62$, $1/\nu=0.89$; the individual expansions give $0.75$, $0.96$ and $0.50$, $0.82$). 
The results are fairly insensitive to the choice of irrelevant corrections: replacing $1/\nu$ by the value of the other expansion at $s=1$ and $1.5$, or both inputs by either endpoint set at $s=1.25$, moves $\Gamma_c$ by at most $3\times10^{-4}$ and the final $z$ by at most $0.001$. 
We take the mean of the three extrapolations as $\Gamma_c$ and half their spread as its uncertainty:
\begin{align}
\Gamma_c&=\GcOne,\ \ \GcTwo,\ \ \GcThree\nonumber\\
&\qquad(s=1,\,1.25,\,1.5).
\label{eq:Gcvalues}
\end{align}
Because the three extrapolated values bracket $\Gamma_c$, the spread is a conservative measure of the extrapolation error.

\begin{table*}[t]
\caption{Finite-size sequences at the three values of $s$. Columns are labeled by the pair $(L,2L)$. $\Gamma_\times$ are the crossings of the structure-factor ratio $R_S$ and the correlation ratio $R_C$ and the self-consistency point $\Gamma^*$, extrapolated with Eq.~\eqref{eq:crossdrift}; $z_{\rm gap}$ and $z_{\Delta_\sigma}$ are evaluated at the extrapolated $\Gamma_c$ of Eq.~\eqref{eq:Gcvalues} and extrapolated with Eq.~\eqref{eq:zeff}. The crossings use the drift exponent $\omega+1/\nu$ and the exponents use $\omega$, both from the theoretical inputs of Sec.~\ref{sec:Gc} ($\omega=0.50$, $0.62$, $0.25$ and $1/\nu=0.83$, $0.89$, $0.91$ at $s=1$, $1.25$, $1.5$); the last column is the fitted amplitude $c$.}
\label{tab:seq}
\begin{ruledtabular}
\begin{tabular}{llccccccc}
$s$ & quantity & (16,32) & (24,48) & (32,64) & (48,96) & (64,128) & extrapolated & $c$\\
\hline
$1.00$ & $\Gamma_\times(R_S)$ & 2.5391 & 2.5307 & 2.5276 & 2.5251 & 2.5242 & 2.5209 & \\
       & $\Gamma_\times(R_C)$ & 2.5217 & 2.5230 & 2.5233 & 2.5233 & 2.5232 & 2.5238 & \\
       & $\Gamma^*$           & 2.5353 & 2.5300 & 2.5277 & 2.5256 & 2.5247 & 2.5227 & \\
       & $z_{\rm gap}$        & 0.5347 & 0.5303 & 0.5279 & 0.5252 & 0.5237 & 0.512 & $+0.088$\\
       & $z_{\Delta_\sigma}$  & 0.4890 & 0.4929 & 0.4953 & 0.4980 & 0.4995 & 0.510 & $-0.085$\\
\hline
$1.25$ & $\Gamma_\times(R_S)$ & 2.0705 & 2.0658 & 2.0641 & 2.0630 & 2.0626 & 2.0612 & \\
       & $\Gamma_\times(R_C)$ & 2.0577 & 2.0600 & 2.0608 & 2.0614 & 2.0617 & 2.0623 & \\
       & $\Gamma^*$           & 2.0602 & 2.0607 & 2.0610 & 2.0614 & 2.0616 & 2.0617 & \\
       & $z_{\rm gap}$        & 0.6524 & 0.6534 & 0.6544 & 0.6565 & 0.6586 & 0.662 & $-0.058$\\
       & $z_{\Delta_\sigma}$  & 0.6610 & 0.6616 & 0.6615 & 0.6610 & 0.6604 & 0.660 & $+0.006$\\
\hline
$1.50$ & $\Gamma_\times(R_S)$ & 1.7705 & 1.7668 & 1.7654 & 1.7643 & 1.7639 & 1.7619 & \\
       & $\Gamma_\times(R_C)$ & 1.7612 & 1.7624 & 1.7628 & 1.7630 & 1.7631 & 1.7637 & \\
       & $\Gamma^*$           & 1.7531 & 1.7567 & 1.7585 & 1.7602 & 1.7610 & 1.7630 & \\
       & $z_{\rm gap}$        & 0.7686 & 0.7746 & 0.7788 & 0.7846 & 0.7889 & 0.837 & $-0.138$\\
       & $z_{\Delta_\sigma}$  & 0.8463 & 0.8446 & 0.8433 & 0.8414 & 0.8399 & 0.825 & $+0.043$\\
\end{tabular}
\end{ruledtabular}
\end{table*}

\subsection{Two independent determinations of $z$}
\label{sec:tworoutes}

\begin{figure*}[t]
\centering
\includegraphics[width=\textwidth]{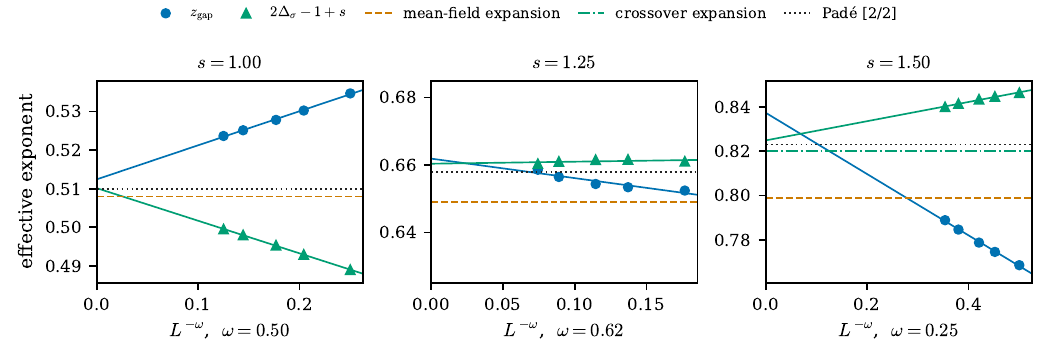}
\caption{The two finite-size estimates of $z$ for the pairs $(L,2L)$, evaluated at the extrapolated $\Gamma_c$, versus $L^{-\omega}$ with $\omega$ fixed to its theoretical input, and the linear fits of Eq.~\eqref{eq:zeff}. Horizontal lines are the mean-field series (dashed), the two-sided Pad\'e (dotted), and, at $s=1.5$ only, where $\delta=1/4$, the crossover expansion (dash-dotted). At $s=1.5$ the gap and static estimators differ by $0.08$ at the pair $(16,32)$ and extrapolate to within $0.012$ of each other; at $s=1.25$ they are within $0.002$ of each other at the pair $(64,128)$ and after extrapolation.}
\label{fig:extrap}
\end{figure*}

At criticality the finite-size gap closes as $\Delta_L\propto L^{-z}$, which gives the dynamical estimate
\begin{equation}
z_{\rm gap}(L,2L;\Gamma)=\log_2\frac{\Delta_L(\Gamma)}{\Delta_{2L}(\Gamma)} .
\label{eq:zgap}
\end{equation}
The second determination uses no energy at all. Under the anisotropic rescaling $x\to bx$, $\tau\to b^z\tau$, the spin transforms as $\sigma\to b^{-\Delta_\sigma}\sigma$, so an equal-time correlator obeys $\langle\sigma(x,0)\sigma(0,0)\rangle=|x|^{-2\Delta_\sigma}$ with no reference to $z$: setting $\tau=0$ leaves nothing to rescale in time. In a ground-state calculation on a ring imaginary time is infinite, the only length is $L$, and $C_L(r)=L^{-2\Delta_\sigma}f(r/L)$, hence $m^2_L\propto L^{-2\Delta_\sigma}$ and
\begin{equation}
z_{\Delta_\sigma}(L,2L;\Gamma)=\log_2\frac{m^2_L(\Gamma)}{m^2_{2L}(\Gamma)}-1+s ,
\label{eq:zD}
\end{equation}
where the exact relation \eqref{eq:shadow} has been used. The two estimators are independent: the gap measures $z$ and nothing about $\Delta_\sigma$ (for $z\neq1$ there is no state--operator correspondence, and the amplitude of $\Delta_L L^z$ is a nonuniversal constant of dimension energy$\times$length$^z$), while the correlator measures $\Delta_\sigma$ and nothing about $z$. 
Their agreement is a nontrivial test of Eq.~\eqref{eq:shadow} on the lattice, which is a prerequisite for the entire crossover construction. 
Both estimators are evaluated at the extrapolated $\Gamma_c$ of Eq.~\eqref{eq:Gcvalues}, the same field for all pairs (Table~\ref{tab:seq}, Fig.~\ref{fig:extrap}). At $s=1.25$ and $1.5$, $z_{\rm gap}$ lies below $z_{\Delta_\sigma}$ at the smallest sizes---by $0.08$ at $s=1.5$---and their difference shrinks monotonically but slowly. Since both estimators are exact in the scaling limit, the discrepancy is a correction to scaling, and the crossover expansion predicts its form: with $\Delta_L=AL^{-z}(1+aL^{-\omega})$,
\begin{equation}
z_{\rm eff}(L,2L)=z+c\,L^{-\omega}+\dots,\qquad c=\frac{a(1-2^{-\omega})}{\ln2} ,
\label{eq:zeff}
\end{equation}
and likewise for the static estimator with its own amplitude. At $s=1.5$ the correction exponent is $\omega=1/4$ (Sec.~\ref{sec:corrections}), so the corrections are of order one at accessible sizes and decay by only a factor $0.84$ per doubling of $L$. This is the regime in which a free power-law fit is unreliable, and it is where the two QMC extrapolations of Ref.~\cite{Shiratani2024} separate. Rather than fitting $\omega$, we fix it to the theoretical input of Sec.~\ref{sec:Gc}, exactly as for the crossings, and fit only $z$ and $c$.
The result is shown in Fig.~\ref{fig:extrap}. 

\subsection{Results and comparison}
\label{sec:results}

Our value of $z$ at each $s$ is the mean of the two extrapolated routes. We estimate its uncertainty by combining in quadrature three contributions: (1), half the difference between the two extrapolated routes, (2), the shift of the mean when $\Gamma_c$ is varied within the bracket \eqref{eq:Gcvalues}, (3), the change when the input $\omega$ in Eq.~\eqref{eq:zeff} is varied by $\pm30\%$ at fixed $\Gamma_c$. 
The three contributions are $(1.1,1.8,0.1)$, $(0.8,1.7,1.3)$, and $(6.2,10.7,8.6)\times10^{-3}$ at $s=1$, $1.25$, $1.5$, giving
\begin{equation}
z=\zOne,\ \ \zTwo,\ \ \zThree\quad(s=1,1.25,1.5).
\label{eq:zvalues}
\end{equation}

\begin{table}[t]
\caption{Critical field and dynamical exponent from DMRG ($L\le\Lmax$), compared with the weak-coupling expansions, the Pad\'e interpolant \eqref{eq:pade}, and the QMC values of Ref.~\cite{Shiratani2024} (power-law and logarithmic extrapolations; the values at $s=1.25$ and $1.5$ are interpolated from their $s=1.2,1.4,1.6$ entries). The QMC critical field at $s=1$ is $2.5264(4)$.}
\label{tab:results}
\begin{ruledtabular}
\begin{tabular}{lccc}
 & $s=1.00$ & $s=1.25$ & $s=1.50$\\
\hline
$\Gamma_c$ (DMRG)            & \GcOne & \GcTwo & \GcThree\\
$z$, DMRG                    & \zOne & \zTwo & \zThree\\
$z$, mean-field series       & 0.508 & 0.649 & 0.799\\
$z$, crossover expansion     & 0.460 & 0.640 & 0.820\\
$z$, Pad\'e, Eq.~\eqref{eq:pade}  & 0.510 & 0.658 & 0.823\\
$z$, QMC power law           & 0.512(2) & $\approx0.655$ & $\approx0.80$\\
$z$, QMC logarithmic         & 0.519(8) & $\approx0.675$ & $\approx0.85$\\
\end{tabular}
\end{ruledtabular}
\end{table}

\begin{figure}[t]
\centering
\includegraphics[width=\columnwidth]{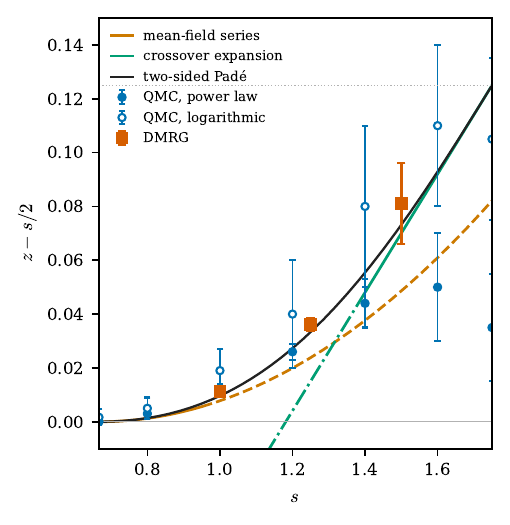}
\caption{Deviation of the dynamical exponent from its Gaussian value $s/2$. Curves: mean-field series (solid where controlled, dashed beyond), crossover expansion (solid near $s_*$, dash-dotted beyond), and the two-sided Pad\'e interpolant \eqref{eq:pade}. Points: QMC power-law (filled) and logarithmic (open) extrapolations of Ref.~\cite{Shiratani2024}, and the present DMRG results, Eq.~\eqref{eq:zvalues} (squares).}
\label{fig:zsummary}
\end{figure}

Table~\ref{tab:results} and Fig.~\ref{fig:zsummary} collect the results. 
At $s=1$ the DMRG value $z=\zOne$ agrees with the mean-field series ($0.508$) and the interpolant ($0.510$) within $0.003$, as it must where $z-s/2$ is still small; the crossover line is far away there ($\delta=0.75$). 
At $s=1.25$, the point farthest from both ends, the DMRG value $z=\zTwo$ lies slightly above the Pad\'e interpolant \eqref{eq:pade}.
At $s=1.5$ the value $z=\zThree$ is $0.011$ above the crossover line $0.820$, Eq.~\eqref{eq:z}, and $0.008$ above the interpolant, both within one standard deviation.
Figure~\ref{fig:zsummary} shows the whole picture in the variable $z-s/2$. We have also included the previous results from QMC for comparison.

\section{Conclusions}
\label{sec:conclusions}

We have proposed a weakly coupled description of the quantum long-range Ising chain near its short-range crossover, in which the equal-time long-range interaction is localized by an instantaneous Gaussian field and the critical theory is a weak perturbation of exactly known Ising CFT data. 

The DMRG calculations test the construction where it is least controlled. Two independent definitions of $z$---dynamical and static, the latter made possible by the exact relation $2\Delta_\sigma=1+z-s$---disagree at accessible sizes near the crossover and are reconciled by the slow $L^{-\omega}$ corrections that the expansions predict, with $\omega=\delta$ near the crossover, after which the dynamical exponent is consistent with the crossover line at $s=1.5$ and with the two-sided Pad\'e interpolation at all three $s$. It should be kept in mind that $\omega$ and $1/\nu$ entered this analysis as theoretical inputs, in the location of the critical point and in the extrapolation alike; the numerical result is therefore a demonstration that the lattice data are consistent with the crossover expansion at $s=1.5$ and with the two-sided interpolation at all three $s$, and not an independent determination of $z$. An analysis with $\omega$ free would require substantially larger systems, because the corrections decay as $L^{-1/4}$ at $s=1.5$. The same mechanism accounts for the systematic difference between power-law and logarithmic QMC extrapolations near $s_*$.

On the analytical side, it is desirable to calculate the critical exponents to higher order in $\delta$.
On the numerical side, it is interesting to measure the dynamical critical exponent $z$ independently using larger system sizes.

A candidate for extending the duality beyond Ising symmetry is the spin-$1/2$ Heisenberg chain studied by Laflorencie, Affleck, and Berciu~\cite{Laflorencie2005}. For an unfrustrated exchange decaying as $1/r^\alpha$, they found a perturbative fixed point for $\alpha$ just above two that separates a phase with long-range N\'eel order from the gapless $SU(2)_1$ conformal phase. The continuum description starts from the level-one $SU(2)$ Wess--Zumino--Witten theory and retains two perturbations: the equal-time long-range interaction between staggered magnetizations $\mathbf n$, and the local marginal current interaction $\mathbf J_L\!\cdot\!\mathbf J_R$. In our construction the first would be represented by coupling $\mathbf n$ to an instantaneous Gaussian vector field. We leave this for future work.
\begin{acknowledgments}
The DMRG calculations were performed with the TeNPy library~\cite{TeNPy}. 
This work was heavily assisted by Claude Fable 5 (Anthropic) and GPT-6 Astra (OpenAI):  the perturbative calculation, numerical codes, the numerical computations reported here, and the drafting of the text were developed in collaboration with them, with the author directing the questions and checking the results.
The work of the author is supported by the European Union (ERC, QFTinAdS, project number 101087025, PI Balt van Rees). Views and opinions expressed are, however, those of the authors only and do not necessarily reflect those of the European Union or the European Research Council Executive Agency.
\end{acknowledgments}

\bibliography{references}

\end{document}